# Identifying the tilt angle and correcting the orbital angular momentum spectrum dispersion of misaligned light beam

*Peng Zhao, Shikang Li, Yu Wang, Xue Feng\*, Kaiyu Cui, Fang Liu, Wei Zhang, and Yidong Huang*

*Department of Electronic Engineering, Tsinghua National Laboratory for Information Science and Technology, Tsinghua University, Beijing, China*

*x-feng@tsinghua.edu.cn*

**Abstract**

The axis tilt of light beam in optical system would introduce the dispersion of orbital angular momentum (OAM) spectrum. To deal with it, a two-step method is proposed and demonstrated. First, the tilt angle of optical axis is identified with a deduced relation between the tilt angle and the variation of OAM topological charges with different reference axes, which is obtained with the help of a charge coupled device (CCD) camera. In our experiments, the precision of measured tilt angle is about $10^{-4}rad$ with topological charges of -3~3. With the measured angle value, the additional phase delay due to axis tilt can be calculated so that the dispersion of OAM spectrum can be corrected with a simple formula while the optical axis is not aligned. The experimental results indicate that the original OAM spectrum has been successfully extracted for not only the pure state but also the superposed OAM states.

**Introduction**

Since OAM was characterized as a new freedom of lightwave by Allen, et.al in 1992 [1], it has been attracted much research interest and shown its potential on various applications. Due to the torque force [2,3] and less radiation pressure deriving from the dark phase singularity [4–6], light beam carrying OAM can serve as optical tweezers and spanners. Meanwhile, as the OAM states can form an infinite dimensional Hilbert space, it can be applied on encoding information for classical optical communications [7,8] or high dimensional quantum entanglement [9,10].

As shown by Allen, *et.al* [1], under paraxial approximation, the angular momentum of light beam can be divided into two independent parts, spin angular momentum (SAM) and OAM. The SAM originates from the circular polarization while the OAM is from the helical wavefront. Usually, Laguerre-Gaussian (LG) mode is utilized to investigate the properties of optical OAM since it is the eigenmode of paraxial wave equation in cylindrical coordinate and any paraxial beam can be decomposed as a series of LG modes. For LG mode with helical phase front of $\exp(-il\phi)$, in which $\phi$ is the azimuthal angle and $l$ is azimuthal index, the OAM in the propagation direction has the discrete value of $l\hbar$ per photon. For such LG mode with nonzero $l$, the intensity distribution is a hollow ring with rotational symmetry by considering light propagation center as pivot. Thus, it is natural to treat this symmetry center as the optical axis of light beam and define the OAM along such axis. However, for real applications, there may be misalignment between the optical axis of incident light beam and that of optical system so that the pure OAM state would transform to the superposition of some OAM states, which is known as the dispersion of OAM spectrum [11,12]. Such misalignment can be introduced by the lateral displacement and (or) the axis tilt between the light beam carrying OAM and optical system. Till now, there are some reports that have investigated such phenomena [11–15] and two methods that have been proposed to deal with it. In Ref. [14], Yi-Dong Liu, et.al have introduced the mean square value of OAM spectrum to quantize the degree of the lateral displacement and the tilt. Their numerical simulation results have shown that the mean square value monotonously increases following the axis misalignment so that a feedback system was proposed to correct it step by step after measuring the mean square value of OAM spectrum. In Ref. 15, J. Lin, *et.al* have deduced the relationship of the dispersed OAM spectrum versus the tilt angle and lateral displacement. Based on such analytical formulation, the misalignment can be extracted from the OAM spectrum, which can be obtained from several coefficients one by one with the help of an optical correlator.

Although both the lateral displacement and the optical axis tilt would introduce OAM spectrum dispersion, the latter is much more critical. This work is focused on this issue and a two-step method is proposed and demonstrated to identify the tilt angle and correct the dispersion of OAM spectrum. First the tilt angle is obtained from the variation of topological charge due to varied reference axes and then the OAM spectrum is corrected according to the tilt angle with a deduced simple formula. Compared with the previously proposed methods, our presented method is relatively simple since there is no additional feedback system [14] or optical correlator [15] required. Actually, only a CCD is required to obtain the Stokes parameter of $S_0$ and the phase distribution of incident light beam. Moreover, the original OAM spectrum of not only a pure OAM state but also a superposition of several OAM states can be well extracted according to our experimental results. At last, since both calculating the tilt angle and correcting the OAM spectrum are based on analytical formulae, these procedures can be automatically and fast done with a computer while there is no iteration required.

**Result Section**

**Principle**

Under the paraxial approximation, the eigenmode of paraxial wave equation in cylindrical coordinate ($r, \phi, z$) is LG mode:

$$LG_p^l = \frac{a_{p,l}}{w}\left(\frac{\sqrt{2}r}{w}\right)^{|l|} L_p^{|l|}\left(\frac{2r^2}{w^2}\right) e^{-r^2/w^2} e^{ikr^2/2R} e^{il\phi}, \quad (1)$$

where $a_{p,l}$ is the normalizing coefficient, $w$ is the width of the light spot, $k$ is the wave vector, $R$ is the curvature radius of the wavefront while $p$ and $l$ denote the radial and azimuthal mode index, respectively. $l$ is also known as the topological charge of the carried OAM since each LG mode of $LG_p^l$ carries OAM of $l\hbar$ and linear momentum of $\hbar k$ per photon along the propagation direction. Theoretically, the electric field of an arbitrary scalar light beam can be expressed as the superposition of LG modes.

$$E(r,\phi,z) = \sum_l \sum_p m_{lp} LG_p^l, \quad (2)$$

where $m_{lp}$ denotes the weight factor of mode $LG_p^l$ and the superposition among the azimuthal index would introduce the OAM spectrum:

$$C_l = \frac{\sum_p |m_{lp}|^2}{\sum_l \sum_p |m_{lp}|^2}. \quad (3)$$

The coefficient of $C_l$ represents the power proportion of $l$-order OAM component over the total power. OAM spectrum plays an important role in many applications, especially in quantum information [9], classic communication [16] and optical imaging [17]. However, it is not easy to measure the OAM spectrum since it relies on the reference axis. In other words, misalignment between the optical axis of incident light beam and that of optical systems would lead to significant distortion on the OAM spectrum. According to Ref. [12] and [13], if there is the tilt or (and) the lateral displacement between the optical axis of the light beam and that of the optical system, the measured spectrum would deviate from the original OAM spectrum carried by the incident beam. Even for a pure OAM state, the measured spectrum would be broadened from a single peak to multiple lines on the OAM spectrum. To quantitatively evaluate the broadening, the dispersion of the OAM spectrum was defined as $\upsilon = \sum_l C_l l^2 - (\sum_l C_l l)^2$ [12,13]. To avoid such dispersion, precise alignment is required while measuring the OAM spectrum. Actually, the axis tilt is much more sensitive than the position displacement on the variation of OAM spectrum. For example, the dispersion value would be about $\upsilon=1$ when the position displacement is about the size of light spot (typically several mm) or the tilt angle is about $2.5\times10^{-4}rad$. Additionally, for state of the art technology, it is easy to correct the position displacement with accuracy of tens of nm, which is much smaller than the typical size of a light spot. But it is still a tough work to correct the tilt angle with accuracy of $\sim10^{-4}rad$. Thus, this work focuses on how to determine the tilt angle and correct the corresponding deviation on OAM spectrum.

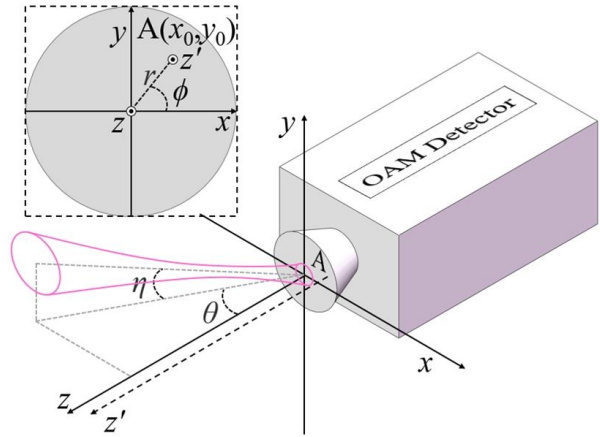

Fig. 1. Schematic of the simplest case that the light beam carrying OAM is incident on the detector with deflection.

Figure 1 shows the simplest example of only a light beam carrying OAM and a detector. The incident light beam is considered as not along the normal direction of the detection plane of the detector. The reference coordinate of the measurement system is considered as that the detection plane is the *x-y* plane while the *z*-axis is the normal direction. Obviously, the *z*-axis is the optical axis of the measurement system to identify the carried topological charge carried by incident beam. Since only the case of an inclined light beam is involved in this work, the origin of the coordinate is settled as the cross point of the incident light beam with *x-y* plane. Thus, as shown in Fig. 1, two angles of $\eta$ and $\theta$ can be employed to describe the direction of incident light beam. The angle of $\eta$ denotes the angle between the optical axis of light beam and its projection on the *x-z* plane, while $\theta$ denotes that between the projection axis and *z*-axis as shown in Fig. 1. It should be mentioned that here the Cartesian coordinate $(x, y)$ is used for simplicity and clarity of demonstration. In latter simulations and experiments, cylindrical coordinate $(r, \phi)$ is adopted for convenience, and the relation between them relies on:

$$\begin{cases} x = r\cos\phi \\ y = r\sin\phi \end{cases}. \quad (4)$$

Actually, if both $\eta$ and $\theta$ are equal to *zero*, the optical axis of the incident light beam is accordant to that of the measurement system so that the value of carried topological charge would be independent to the choice of *z*-axis. Specifically, if *z*-axis is moved to a parallel axis of *z'*-axis, which is defined by the point of $A(x_0, y_0)$ in *x-y* plane as shown in Fig. 1, the measured topological charge of incident light beam would be as constant value due to the intrinsic nature of OAM. However, extrinsic OAM would be introduced if there is a tilt angle between the optical axis of the incident light beam and measurement system (if anyone of $\eta$ and $\theta$ is not equal to *zero*). Here, $l_z$ and $l'_z$ denote the topological charge defined by the reference axis as *z*-axis and *z'*-axis, respectively. According to Ref. [13], the relation between $l_z$ and $l'_z$ is:

$$l'_z = l_z + \frac{-x_0 P_y + y_0 P_x}{\hbar}, \quad (5)$$

where $P_x$ and $P_y$ denotes the *x* component and *y* component of transversal linear momentum of incident light beam, respectively. In this work, the angle $\eta$ and $\theta$ are assumed small enough so that small-angle approximation of $\sin\eta \approx \eta$ and $\sin\theta \approx \theta$ can be adopted. Hence, the values of linear momentum are $P_x = \hbar k\theta$ per photon and $P_y = -\hbar k\eta$. Consequently, the variation between $l_z$ and $l'_z$ can be deduced from Eq. (5) as:

$$\Delta l = l'_z - l_z = \frac{-x_0 P_y + y_0 P_x}{\hbar} = x_0 k\eta + y_0 k\theta. \quad (6)$$

According to Eq. (6), if the incident light beam is well aligned with the optical axis of measurement system ($\eta=0$ and $\theta=0$), there is no extrinsic OAM ($l_z = l'_z$). However, if there is a tilt angle ($\eta \neq 0$ or $\theta \neq 0$), extrinsic OAM ($l=l'_z - l_z$) would be introduced and the tilt angle of $\eta$ and $\theta$ can be extracted as:

$$\begin{cases} \eta = \frac{1}{k}\frac{\partial l}{\partial x} \\ \theta = \frac{1}{k}\frac{\partial l}{\partial y} \end{cases}. \quad (7)$$

Equation (7) manifests that relation of the tilt angles and the topological charge variation along the coordinate axis. Specifically, the angle of $\eta$ and $\theta$ are proportional to the gradient of topological charge over the *x*- and *y*-axis, respectively. Actually, Eq. (7) also provides a simple method to obtain the angles of $\eta$ and $\theta$ by measuring topological charge with different reference axis. After the angles of $\eta$ and $\theta$ are known, the most direct method to correct the tilt is precisely adjusting the optical system. For some applications, such precise alignment is inevitable. For example, for optical wireless communication with OAM multiplexing [16], the optical axis of transmitter and receiver has to be aligned to avoid channel cross talk since different OAM mode serves as different transmission channel. However, for some applications where only the actual OAM spectrum is required, another option is to extract the OAM spectrum with the known tilt angles of $\eta$ and $\theta$. In the next section, a two-step method to measure tilt angles and correct the measured OAM spectrum will be demonstrated based on our previous work [18].

For the sake of simplicity, the experiments and discussions would focus on a simplified situation of $\eta=0$. In another word, the light beam carrying OAM is incident in the *x-z* plane as shown in Fig. 2. Here, the detection plane is *x-y* plane while the green plane in Fig. 2(a) is the transverse plane normal to the optical axis of incident light beam, which is inclined to the *x-y* plane with angle of $\theta$ as shown in Fig. 2(b).

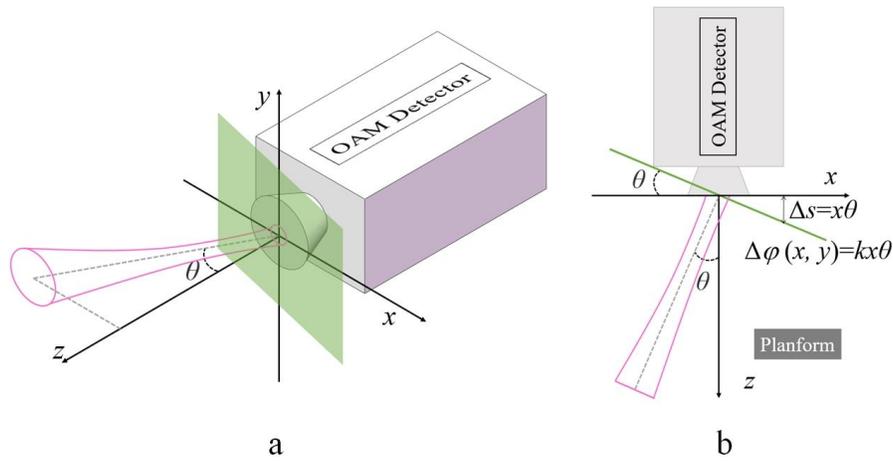

Fig. 2. The schematic of (a) the 3-D view and (b) top view along *y*-axis of the simplified case that the incident light beam carrying OAM is in the *x-z* plane ($\eta=0$).

The first step is to obtain the tilt angle of $\theta$. The topological charges with different reference axis should be measured according to Eq. (7). Theoretically, only two different reference axes normal to the detection plane are required to measure $\Delta l$. However, to reduce the fluctuations during measuring process, multiple reference axes can be adopted to calculate the gradient in Eq.(7) and the number is 21×21 in our experiments as shown in the Results section. Then the tilt angle can be obtained by linear fitting the measured topological charges with a number of different reference axes. In our previous work [18], the topological charge of arbitrary vectorial light beam can be expressed with Stokes parameters and Pancharatnam phase of the transverse field of the light beam. For uniform polarized light beam considered here, the formula can be reduced as:

$$l = \frac{\iint -S_0 \frac{\partial \varphi}{\partial \phi} r dr d\phi}{\iint S_0 r dr d\phi}, \quad (8)$$

where $S_0$, $l$, and $\varphi$ denote Stokes parameters, the total topological charge, and the phase distribution of light beam, respectively, while $r$ and $\phi$ denote the radial and azimuthal coordinates of cylindrical coordinate system, respectively, sharing the same origin and $z$-axis with the Cartesian coordinate system as shown in Fig. 1. Stokes parameters and the phase distribution can be obtained by measuring the intensities of light beam and the interference of light beam with one reference light, respectively. With Eq. (8), the topological charges with different reference axes can be obtained by moving the position of the coordinate origin. Since the topological charges are obtained from the gradient of spiral spatial phase and the gradient would change when moving the position of the coordinate origin.

The second step is extracting the measured spectrum. With Stokes parameters and the phase distribution, the OAM spectrum of light can be deduced as [19]:

$$C_l = \frac{\int \left| \int \sqrt{S_0} e^{-i\varphi} \frac{e^{-il\phi}}{\sqrt{2\pi}} d\phi \right|^2 rdr}{\iint S_0 rdrd\phi}. \quad (9)$$

As shown in Fig. 2, the Stokes parameter ($S_0$) and the phase distribution ($\varphi$) is measured on the $x$-$y$ plane (the detection plane), but these parameters should be measured on the exact transverse plane of incident light beam so that the original OAM spectrum carried by the incident light beam can be extracted. Under small axis tilt, the measured Stokes parameters could be treated a constant value. For the phase distribution, there is an additional phase delay due to deviation between the transverse plane of incident light beam and the detection plane. As shown in Fig. 2(b), the extra propagation length is approximated as $\Delta s(x, y) = \theta x$ so that additional phase delay is $\Delta \varphi (x, y) = kx\theta$. Hence, the original phase distribution can be obtained as $\varphi = \varphi_m - \Delta \varphi$ from the measured phase distribution of $\varphi_m$.

**Experimental setup:**

To verify our proposed method of measuring the tilt angle and correcting the dispersion of the OAM spectrum, a series of experiments have been carried out.

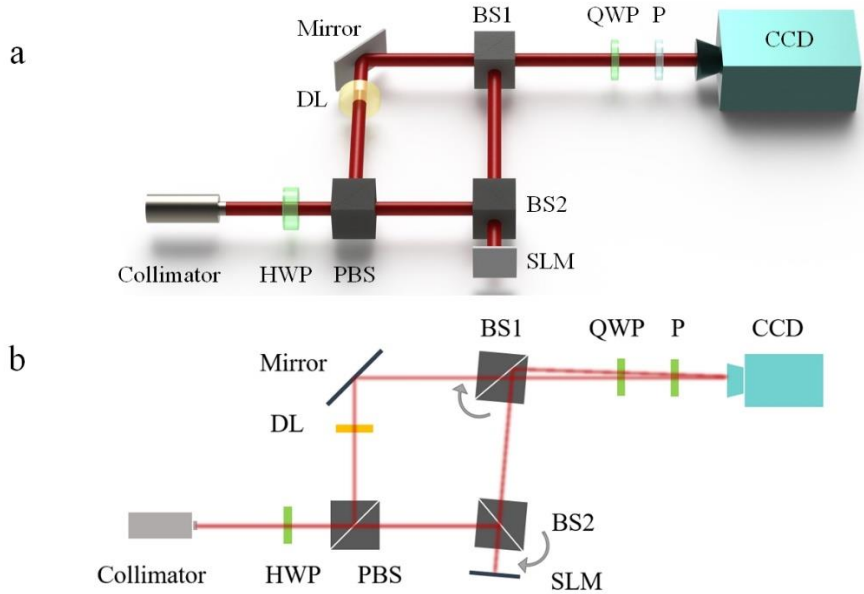

Fig. 3. (a) The schematic of experimental arrangements. HWP: half-wave plate, PBS: polarizing beam splitter, BS: beam splitter, SLM: spatial light modulator, DL: delay line, QWP: quarter-wave plate, P: polarizer. (b) The tilt angle between the optical axis of the objective and reference light beam is controlled by rotating BS1, BS2 and SLM.

Figure 3(a) shows the schematic of experimental setup, which is a typical Mach-Zender interferometer arrangement. A laser operating at wavelength of 1550nm (RIO Orion) is connected with a collimator by a pigtail fiber so that the diameter of light beam is expanded to several millimeters and injected into the optical system. Followed the collimator, there are a half-wave plate (HWP) and a polarizing beam splitter cube (PBS) to split light beam into two orthogonal polarization components with controlled ratio. The horizontal polarization component is incident towards a beam splitter of BS2 and reflected to the spatial light modulator (SLM, PLUTO-TELCO-013). Then a pre-settled OAM mode would be generated through the SLM by modulating the phase of incident light, which is treated as the objective light. The vertical polarization component reflected by PBS would serve as the reference light for interference. Then the objective and the reference light beams would be combined by BS1 and a delay line (DL) is inserted within the optical path of reference light to control the phase delay. After BS2, the interference pattern would be detected by a CCD, in front of which there are a quarter-wave plate (QWP) and a polarizer (P) to filter out the right-hand circularly polarized component. The detected intensity pattern would be recorded and processed by a computer connected to the CCD. First, the whole system is well aligned as Fig. 3(a), in which both the optical axes of the objective beam and reference beam are accordant and normal to the detection plane of CCD. To simulate the light beam carrying OAM incident on the detector with a small tilt angle, the BS1, BS2 and SLM are rotated slightly. As shown in Fig. 3(b), with simultaneously rotating the BS2 and the SLM, the objective and reference light beam would be incident on the different position on the reflection plane of the BS1. Next, by rotating the BS1, the objective beam and the reference beam can be aligned at the same position on the detection plane of CCD camera. Although the BS2 and the SLM as well as BS1 have been rotated, the propagation path of the reference beam is not changed. Thus, the reference coordinate of measuring system can be constructed by the optical axis of reference beam ($z$-axis) and the detection plane of CCD camera ($x$-$y$ plane). In principle, the tilt angle between the optical axis of incident beam and the reference coordinate of measuring system can be obtained by reading the rotating angle of BS1. However, the tilt angle would be too small to obtain from the graduation on the rotation stage in our experiment. Thus, the accurate tilt angle is measured by counting the interference pattern shift of two Gaussian light beams and the detailed method can be found in the supplementary material. Moreover, the measured tilt angle with such method is recorded as standard angle and would be compared with that obtained by measuring the extrinsic topological charge.

As mentioned, Stokes parameter of $S_0$ and the phase distribution of objective light should be measured. The Stokes parameter of $S_0$ is related to the intensity of objective light, which can be directly obtained from CCD camera. To measure the phase distribution of objective light, we have measured the interference pattern between the objective and reference light. As shown in Fig. 3, the objective light is OAM beam while the reference light is 0-order LG mode. Here, we denote the intensity and phase distribution of objective/reference beams on the plane of CCD camera as $I_{obj}/I_{ref}$ and $\varphi_{obj}/\varphi_{ref}$, respectively. Then, the Stokes parameter is $S_0=I_{obj}$, and the interference pattern ($I_{inter}$) of such two beams can be deduced as:

$$I_{inter}(r,\phi) = I_{obj}(r,\phi) + I_{ref}(r,\phi) + 2\sqrt{I_{obj}(r,\phi)I_{ref}(r,\phi)}\cos(\varphi_{obj}(r,\phi) - \varphi_{ref}(r,\phi) - \varphi_{del}), \quad (10)$$

where $\varphi_{del}$ represents the extra phase delay in reference beam and can be adjusted by the delay line (DL in Fig. 3). Since the reference light beam is the 0-order LG mode, the phase distribution of reference light satisfies $\varphi_{ref}(r,\phi) \propto r^2$ in the condition that the optical axis of the reference light beam coincides with the optical axis of detector. For LG mode, the phase factor is proportional to $r^2$ and $e^{ikr^2/2R}$ determines the radius of the curvature of the wavefront. Thus, the phase term of $\varphi_{obj}(r,\phi)-\varphi_{ref}(r,\phi)$ is equivalent to the phase of a light beam with the same features in terms of Stokes parameters, the topological charge and the direction of propagation, as the objective light beam only except for the curvature radius of the wavefront. And the topological charge of objective light beam can be calculated after substituting the $\varphi_{obj}(r,\phi)-\varphi_{ref}(r,\phi)$ into Eq. (7) as the $\varphi$. In principle, two different but arbitrary values of $\varphi_{del}$ are needed to extract the phase distribution of $\varphi_{obj}(r, \phi)$-$\varphi_{ref}(r, \phi)$. Here, two values of $\varphi_{del1}$ and $\varphi_{del2}$ with the relation of $\varphi_{del2}=\varphi_{del1}+\pi/2$ are adopted to achieve the maximum signal-to-noise ratio of the calculated phase distribution. By recording the interference patterns as $I_{inter1}$ and $I_{inter2}$ with corresponding phase delay of $\varphi_{del1}$ and $\varphi_{del2}$, respectively, the phase distribution can be obtained as:

$$\varphi_{obj}(r,\phi) - \varphi_{ref}(r,\phi) = \arg(I_{inter1}(r,\phi) - I_{obj}(r,\phi) - I_{ref}(r,\phi), I_{inter2}(r,\phi) - I_{obj}(r,\phi) - I_{ref}(r,\phi)) \quad (11)$$

The detailed deduction from Eq. (10) to Eq. (11) can be found in the supplementary material.

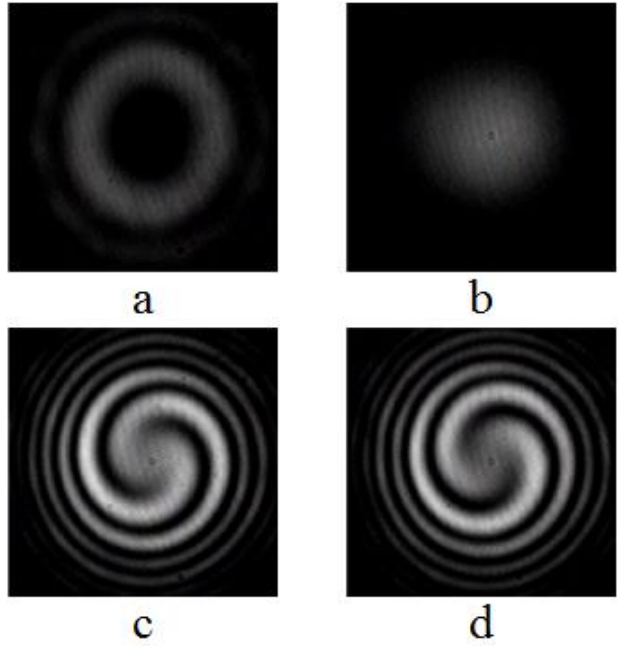

Fig. 4. (a) $I_{obj}(r, \phi)$: the intensity of pre-settled OAM mode beam (objective beam), (b) $I_{ref}(r, \phi)$: the intensity of reference light beam, (c) $I_{inter1}(r, \phi)$ and (d) $I_{inter2}(r, \phi)$: interference intensities of objective beam and reference beam.

As an example, Fig. 4 shows the measured results obtained with the objective beam carrying OAM of $l=-2$. Figure 4(a) and (b) are the intensity patterns of the objective and reference beams ($I_{obj}(r, \phi)$ and $I_{ref}(r, \phi)$), while Fig. 4(c) and (d) are the interference patterns of $I_{inter1}(r, \phi)$ and $I_{inter2}(r, \phi)$. In both Fig. 4(c) and (d), there are two petals that coincide with the objective beam carrying OAM of $l=-2$. Moreover, the interference pattern of Fig. 4(d) rotates half a quarter circumference compared with Fig. 4(c), which is also accordant to additional phase delay of $\pi/2$ in Fig. 4(d).

**Results:**

## Measure the Tilt Angle

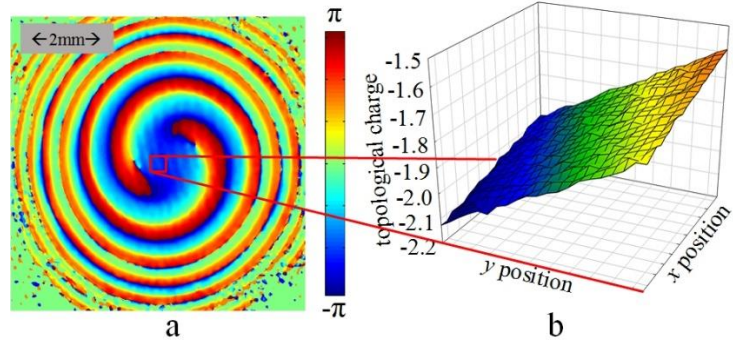

Fig. 5. (a) The phase distribution of the objective beam carrying OAM of $l=-2$, and the red box section denotes the boundary of the reference axes adopted to calculate the topological charges. (b) The topological charges calculated with different reference axes (21×21 axes are chosen to reduce the fluctuations) and the *x-y* coordinate denotes the position of reference axes while the vertical coordinate denotes the corresponding topological charge.

As mentioned in Principle section, to obtain the tilt angle between the optical axis of objective light beam and that of reference beam, the topological charge with different reference axes should be calculated. The results are shown in Fig. 5. Figure 5(a) shows the phase distribution of the objective beam carrying OAM of $l=-2$ and the gray marker shows the plotting scale. In principle, any point within *x-y* plane can be settled as the coordinate origin so that the corresponding *z'*-axis can be determined and then the corresponding topological charge can be obtained in success. In our experiment, the 21×21 pixels around the center of detection plane have been settled as the coordinate origin, which is shown as the red box in Fig. 5(a). Thus, 21×21 reference axes as well as the values of topological charge can be obtained and the corresponding results are summarized in Fig. 5(b). In Fig. 5(b), it can be seen that there is a gradient of topological charge along the *y*-axis. The reason is that the objective beam is incident in the *x-z* plane ($\eta=0$). If both tilt angles of $\eta$ and $\theta$ are none-zero, there would be the gradient of topological charge along both *x*- and *y*-axis according to Eq. (7). By linear fitting the obtained topological charge versus the *y*-axis in Fig. 5(b), the tilt angle can be calculated as of $\theta = -3.37\times10^{-4} rad$.

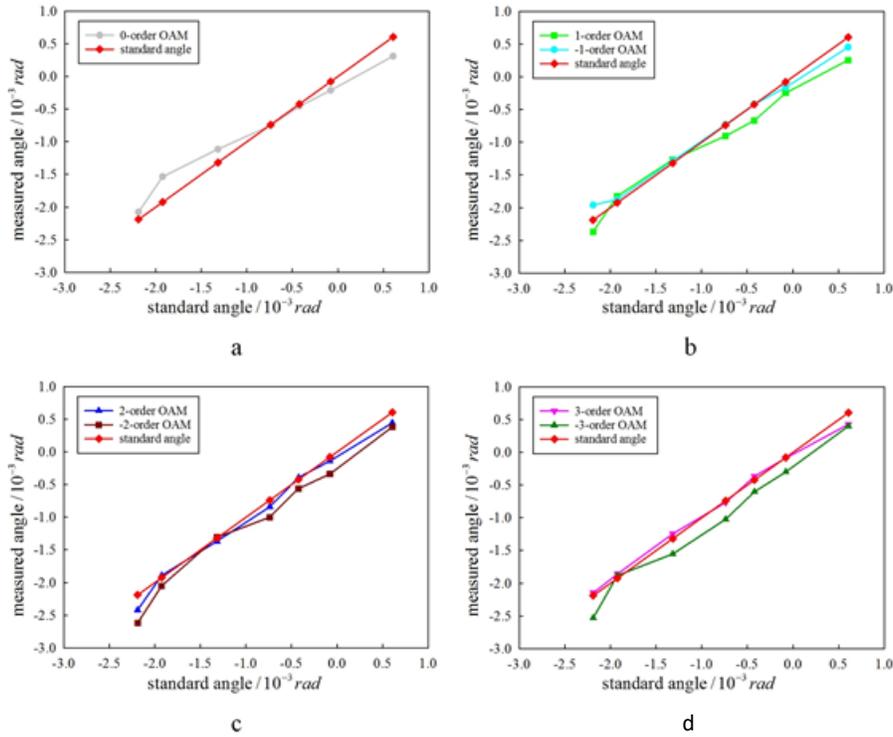

Fig. 6. The measured tilt angle, with beams carrying OAM of $l$=-3~3. The horizontal coordinate denotes the standard tilt angles obtained by the pattern shifts of the interference of two Gaussian beams while the standard tilt angle is plotted as a diamond line with slope of 1. (a) $l$=0. (b) $l$=-1 and 1. (c) $l$=-2 and 2. (d) $l$=-3 and 3.

Furthermore, a series of experiments have been carried out with the objective beam carrying OAM of $l$=-3~3 and varied tilt angle. As mentioned above, the tilt angle is first measured by interference pattern shift of two Gaussian beams and recorded as standard angle as reference for comparison with the values measured by our proposed method. Figure 6 shows the summarized results. In each Figure, the horizontal axis is the standard angle and the standard tilt angle is also plotted as a hollow diamond line with slope of 1 for clear comparison. From Fig. 6, it could be found that the tilt angles measured by our method are well consistent with the standard values and the precision in our system is about $10^{-4}$ $rad$.

**Correct the OAM spectrum**

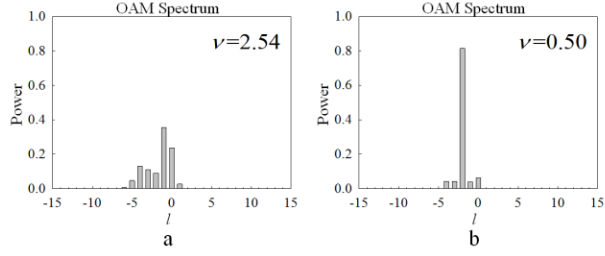

Fig. 7. (a) and (b) shows the results of the measured OAM spectrum, and the corrected OAM spectrum with objective beam carrying OAM of $l=-2$ and tilt angle of $-3.37\times10^{-4} rad$, and the OAM spectrum dispersion factor of $\upsilon = \sum_l C_l l^2 - (\sum_l C_l l)^2$ is also presented.

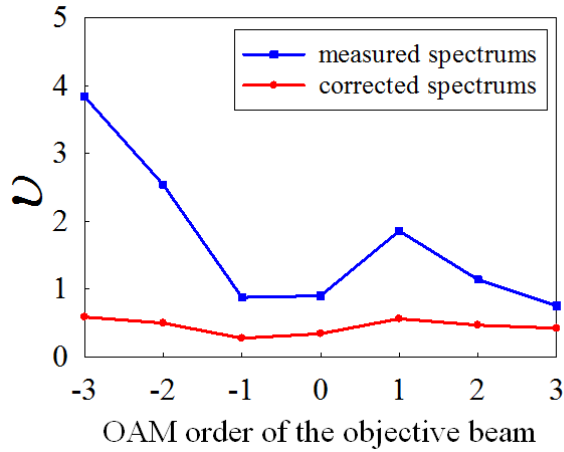

Fig. 8. The dispersion factor of measured OAM spectrums and corrected OAM spectrums. The objective beams carries OAM of $l=-3\sim3$.

After measuring the tilt angle, the next step is correcting the OAM spectrum by correcting the measured phase distribution as mentioned above. A typical result is shown in Fig. 7. The pre-settled objective light beam carries OAM of $l=-2$, which is the same as the sample in Fig. 4 and Fig. 5. With the tilt angle of $-3.37\times10^{-4} rad$, the measured spectrum (Fig. 7(a)) shows multiple lines on OAM spectrum instead of one single peak in the position of $l=-2$. Figure 7(b) shows the corrected OAM spectrum with our proposed algorithm. It can be seen that the OAM spectrum is significantly improved. In order to quantitively evaluate the correction accuracy, the OAM spectrum dispersion factor of $\upsilon = \sum_l C_l l^2 - (\sum_l C_l l)^2$ is calculated. Obviously, if the dispersion factor is $\upsilon=0$, it means that there is no degradation. In Fig. 7, the value is $\upsilon=2.54$ for the directly measured OAM spectrum while that is improved to $\upsilon=0.50$ for the corrected OAM spectrum. Although the dispersion factor did not achieve the ideal value of $\upsilon=0$, the improvement is very remarkable. The OAM spectra with and without correction for the objective beam of $l=$-3~3 are shown in the supplementary material. Here, Fig. 8 shows the summarized result of the dispersion factor $\upsilon$ of measured and corrected OAM spectra as a comparison. It can be found that the dispersion factor is improved to $\upsilon$=0.3~0.6 with our correction method.

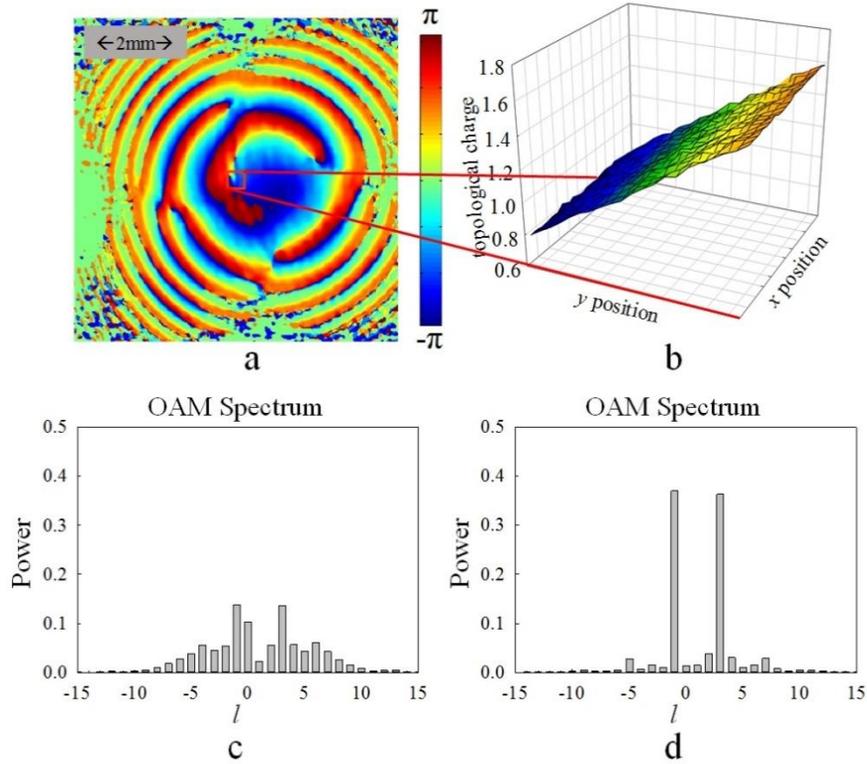

Fig. 9. The experimental results for light beam superposed by OAM of $l$=3 and -1 with power ratio of 50:50. (a) The intensity of objective beam, and the red shadow section denotes boundary of the reference axes used to calculate the topological charges. (b) The topological charges calculated with different reference axes and the horizontal coordinate denotes the position of reference axes while the vertical coordinate denotes the corresponding topological charge. (c) Directly measured OAM spectrum. (d) Corrected OAM spectrum.

Till now, the objective beam only carries a pure OAM state. Actually, our method can also be applied on the light beams with superposition OAM states. As a concrete example, the superposition of $l$=3 and $l$=-1 is settled as the objective beam. Obviously, each of OAM components has the same tilt angle, since they are generated simultaneously by the SLM. Thus the average topological charge of the superposition state would also be affected as the pure OAM state and can be employed to measure the tilt angle. In experiment, the power ratio between 3-order and -1-order component is settled as 50:50 and the measured light field is shown in Fig. 9(a) while Fig. 9(b) is the calculated average topological charges with different reference axes. From the result of Fig. 9(b), the tilt angle can be obtained as $-6.15\times10^{-4}$ *rad*. Figure 9(c) and (d) shows the directly measured OAM spectrum and the corrected OAM spectrum, respectively. It could be seen that the OAM spectrum has been well corrected for the light beam carrying superposition of two OAM modes. More results for beams carrying the superposed OAM states are also shown in supplementary material.

**Discussion:**

Till now, our method to deal with inclined optical axis of light beam carrying OAM has been demonstrated. The whole procedure could be summarized as: First, the field intensity distribution of objective light beam is detected with a CCD camera. Further, by measuring two interference patterns of objective beam and a Gaussian beam, the phase distribution of the objective light filed is obtained so that the topological charges with different reference axes can be calculated. After that, the tilt angle can be obtained from the gradient of topological charge along the coordinate axis. Once the tilt angle is identified, one option is aligning the optical axis according to it and the other is applying the second-step of our proposed method to correct the OAM spectrum. For some applications, the received light beam or photons would be processed further, *e.g.* optical wireless communication or quantum entanglement, it is inevitable to adjust the optical system since the OAM carried by each photon has truly changed. However, if there is no further manipulation on the received light beam and only the original OAM spectrum is required, there is no necessary to adjust the optical system since our proposed method does not rely on the alignment of optical axis.

Additionally, it should be mentioned that the key point of our method is to obtain the tilt angle from the topological charge diversity with different reference axes. But the result is independent to the method of measuring the topological charge so that one can employ any methods to obtain it, *e.g.* OAM hologram and optical transformation elements. Here, the reason of measuring the field distribution of light beam in this work is only due to the applicability of any light beam and convenience of implementation. Moreover, the interference method to obtain the phase distribution of objective beam is also replaceable and any other method, *e.g.* computational optical method, can be adopted.

At last, although the LG modes are chosen as the objective beams in this work, it should be mentioned that there is a similar analysis and relation between the tilt angle and the topological charge diversity for any light beam and obviously, such relation would be more complicated. Thus, our proposed the method can be extended for any light beams with similar procedure but modified calculation and correction formulae.

In this paper, the OAM topological charges of light beam with different reference axes have been deduced and a two-step method to identify the tilt angle and correct the OAM spectrum of misaligned light beam without iteration is proposed and demonstrated in success. In our experiment, the field intensity and phase distribution of light beam are measured with a CCD camera so that the tilt angle of optical axis can be obtained. With employing light beams carrying OAM of $l$= -3~3 as the objective beams, the precision of measuring tilt angles is about $10^{-4} rad$. After that, the OAM spectrum can be corrected without adjusting the optical axis and for pure OAM state beams, the dispersion factors can be improved from 0.8~3.9 to 0.3~0.6 with our correction algorithm. Moreover, the spectrum correction is also demonstrated with beams of superposed OAM states.

**References**
1. Allen, L., Beijersbergen, M. W., Spreeuw, R. J. C. & Woerdman, J. P. Orbital angular momentum of light and the transformation of Laguerre-Gaussian laser modes. *Phys. Rev. A* **45,** 8185 (1992).
2. T O'Neil, A. & Padgett, M. J. Three-dimensional optical confinement of micron-sized metal particles and the decoupling of the spin and orbital angular momentum within an optical spanner. *Opt. Commun.* **185,** 139–143 (2000).
3. He, H., Friese, M. E. J., Heckenberg, N. R. & Rubinsztein-Dunlop, H. Direct observation of transfer of angular momentum to absorptive particles from a laser beam with a phase singularity. *Phys. Rev. Lett.* **75,** 826 (1995).
4. Ashkin, A. Forces of a single-beam gradient laser trap on a dielectric sphere in the ray optics regime. *Biophys. J.* **61,** 569 (1992).
5. Simpson, N. B., McGloin, D., Dholakia, K., Allen, L. & Padgett, M. J. Optical tweezers with increased axial trapping efficiency. *J. Mod. Opt.* **45,** 1943–1949 (1998).
6. O'Neil, A. T. & Padgett, M. J. Axial and lateral trapping efficiency of Laguerre–Gaussian modes in inverted optical tweezers. *Opt. Commun.* **193,** 45–50 (2001).
7. Wang, J. *et al.* Terabit free-space data transmission employing orbital angular momentum multiplexing. *Nat. Photonics* **6,** 488–496 (2012).
8. Zhang, D., Feng, X. & Huang, Y. Encoding and decoding of orbital angular momentum for wireless optical interconnects on chip. *Opt. Express* **20,** 26986–26995 (2012).

9. Vaziri, A. *Entanglement of the orbital angular momentum states of photons*. (na, 2001).

10. Vaziri, A., Weihs, G. & Zeilinger, A. Experimental Two-Photon, Three-Dimensional Entanglement for Quantum Communication. *Phys. Rev. Lett.* **89,** (2002).

11. Lavery, M. P., Berkhout, G. C., Courtial, J. & Padgett, M. J. Measurement of the light orbital angular momentum spectrum using an optical geometric transformation. *J. Opt.* **13,** 064006 (2011).

12. Vasnetsov, M. V., Pas'ko, V. A. & Soskin, M. S. Analysis of orbital angular momentum of a misaligned optical beam. *New J. Phys.* **7,** 46–46 (2005).

13. Zambrini, R. & Barnett, S. M. Quasi-Intrinsic Angular Momentum and the Measurement of Its Spectrum. *Phys. Rev. Lett.* **96,** (2006).

14. Liu, Y.-D., Gao, C., Qi, X. & Weber, H. Orbital angular momentum (OAM) spectrum correction in free space optical communication. *Opt. Express* **16,** 7091–7101 (2008).

15. Lin, J., Yuan, X.-C., Chen, M. & Dainty, J. C. Application of orbital angular momentum to simultaneous determination of tilt and lateral displacement of a misaligned laser beam. *JOSA A* **27,** 2337–2343 (2010).

16. Gibson II, G. *et al.* Increasing the data density of free-space optical communications using orbital angular momentum. in (eds. Ricklin, J. C. & Voelz, D. G.) 367 (2004). doi:10.1117/12.557176

17. Uribe-Patarroyo, N., Fraine, A., Simon, D. S., Minaeva, O. & Sergienko, A. V. Object identification using correlated orbital angular momentum states. *Phys. Rev. Lett.* **110,** 043601 (2013).

18. Zhang, D., Feng, X., Cui, K., Liu, F. & Huang, Y. Identifying Orbital Angular Momentum of Vectorial Vortices with Pancharatnam Phase and Stokes Parameters. *Sci. Rep.* **5,** 11982 (2015).

19. Molina-Terriza, G., Torres, J. P. & Torner, L. Management of the Angular Momentum of Light: Preparation of Photons in Multidimensional Vector States of Angular Momentum. *Phys. Rev. Lett.* **88,** (2001).

**Acknowledgments**

This work was supported by the National Natural Science Foundation of China (Grant No. 61675112 and 61321004) and the National Basic Research Program of China (No. 2013CBA01704 and No. 2013CB328704).

**Author Contributions**

P.Z., S.L. and X.F. developed the theory. P.Z. and S.L. ran the numerical simulations. P.Z., S.L. and Y.W. did the experiments. P.Z. and X.F. wrote the paper. K.C., F.L. and W.Z. provided useful discussions. Y.H. revised the manuscript. All authors read and approved the manuscript.

# Supplementary Information

# Identifying the tilt angle and correcting the orbital angular momentum spectrum dispersion of misaligned light beam

*Peng Zhao, Shikang Li, Yu Wang, Xue Feng\*,Kaiyu Cui, Fang Liu, Wei Zhang, and Yidong Huang*

*Department of Electronic Engineering, Tsinghua National Laboratory for Information Science and Technology, Tsinghua University, Beijing, China*

[x-feng@tsinghua.edu.cn](x-feng@tsinghua.edu.cn)

**Detailed Deduction of the equations:**
The deduction of the Eq. (11) from Eq. (10) in the main text:

$$I_{inter}(r,\phi) = I_{obj}(r,\phi) + I_{ref}(r,\phi) + 2\sqrt{I_{obj}(r,\phi)I_{ref}(r,\phi)}\cos(\varphi_{obj}(r,\phi) - \varphi_{ref}(r,\phi) - \varphi_{del})$$

Putting in the $\varphi_{del1}$ and $\varphi_{de2}$, with the relation of $\varphi_{del2} = \varphi_{del1} + \pi/2$, the interference patterns, denoted as $I_{inter1}$ and $I_{inter2}$ can be deduced as:

$$I_{inter1}(r,\phi) = I_{obj}(r,\phi) + I_{ref}(r,\phi) + 2\sqrt{I_{obj}(r,\phi)I_{ref}(r,\phi)}\cos(\varphi_{obj}(r,\phi) - \varphi_{ref}(r,\phi) - \varphi_{del1})$$

$$I_{inter2}(r,\phi) = I_{obj}(r,\phi) + I_{ref}(r,\phi) + 2\sqrt{I_{obj}(r,\phi)I_{ref}(r,\phi)}\cos(\varphi_{obj}(r,\phi) - \varphi_{ref}(r,\phi) - \varphi_{del2})$$

$$= I_{obj}(r,\phi) + I_{ref}(r,\phi) + 2\sqrt{I_{obj}(r,\phi)I_{ref}(r,\phi)}\sin(\varphi_{obj}(r,\phi) - \varphi_{ref}(r,\phi) - \varphi_{del1})$$

Then:

$$2\sqrt{I_{obj}(r,\phi)I_{ref}(r,\phi)}\cos(\varphi_{obj}(r,\phi) - \varphi_{ref}(r,\phi) - \varphi_{del1}) = I_{inter1}(r,\phi) - I_{obj}(r,\phi) - I_{ref}(r,\phi)$$

$$2\sqrt{I_{obj}(r,\phi)I_{ref}(r,\phi)}\sin(\varphi_{obj}(r,\phi) - \varphi_{ref}(r,\phi) - \varphi_{del1}) = I_{inter2}(r,\phi) - I_{obj}(r,\phi) - I_{ref}(r,\phi)$$

Since the phase delay of $\varphi_{del1}$ is constant, ignore the term without loss the generality, the phase difference between the object light beam and the reference light beam can be deduced as the Eq. (10):

$$\varphi_{obj}(r,\phi) - \varphi_{ref}(r,\phi) = \arg(I_{inter1}(r,\phi) - I_{obj}(r,\phi) - I_{ref}(r,\phi), I_{inter2}(r,\phi) - I_{obj}(r,\phi) - I_{ref}(r,\phi))$$

**Measurement of the tilt angle:**
Under the Cartesian coordinates $(x, y, z)$, the light field of the Gaussian beam propagating along the $z$-axis can be expressed as:

$$E(x,y,z) = E_0 \frac{\omega_0}{\omega(z)} \exp\left(\frac{-(x^2+y^2)}{\omega(z)^2}\right) \exp\left(ik\frac{(x^2+y^2)}{2R(z)}\right) \exp(ikz),$$

where $E_0$ denotes normalized coefficient of the field, while $\omega_0$ and $\omega(z)$ denotes the size of the light waist and light spot at the position of $z$, respectively. $R(z)$ and k denotes the curvature radius of the wavefront at $z$ and the wave vector.

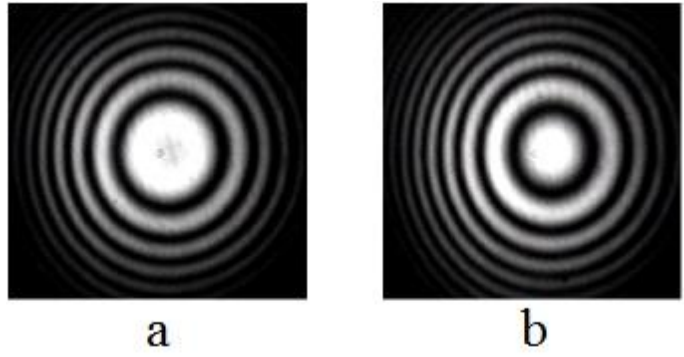

Fig. 1. The interference patterns of two Gaussian beams, while there is without tilt and with tilt between the two Gaussian beams in (a) and (b).

Considering two Gaussian beams with the likely intensity distribution but different phase distribution, from different $R(z)$. The dark fringe positions in the interference pattern of such two Gaussian beams is mainly the positions of the phase difference between two beams is odd times of π. Thus the position function of the dark fringes can be expressed as:

$$\frac{1}{2}k\left(\frac{x^2+y^2}{R_1(z)}-\frac{x^2+y^2}{R_2(z)}\right)=(2n+1)\pi,$$

where $R_1(z)$ and $R_2(z)$ denotes the curvature radiuses of the wavefront at $z$ of the two Gaussian beams. Determined by the equation, the dark fringes are several concentric rings with the center at position $(x, y)=(0, 0)$, as shown in Fig. 1(a). The coefficient of $\frac{1}{R_1(z)}-\frac{1}{R_2(z)}$ determines the intervals between the rings. Thus the coefficient could be extracted from the distribution the rings. In our experiments, the two Gaussian beams come from objective light and reference light, respectively. Considering the objective light having a tilt between reference light as the situation shown in the Fig. X in the main body, the phase distribution of the objective light would add an extra phase $\Delta\varphi(x, y)=k\theta x$, according to the analysis in the last part of the introduction section. The dark fringes position function of the interference pattern could be deduced as:

$$\frac{1}{2}k\left(\frac{x^2+y^2}{R_1(z)}+\theta x-\frac{x^2+y^2}{R_2(z)}\right)=(2n+1)\pi.$$

Thus, the center of the concentric rings shifts from 0 to $-\frac{\theta}{2\left(\frac{1}{R_1(z)}-\frac{1}{R_2(z)}\right)}$, while the intervals between the rings remains, as shown in Fig. 1(b). The tilt angle of the objective light beam can be extracted from such position shift.

To demonstrate the generality of our method, pure OAM state beams of -3~3-order and several superposition OAM beams are adopted as the objective light beam, respectively. Here we show the results of OAM spectrum correction.

**Pure state OAM beams:**

Figure 3 shows the measured OAM spectrums and the corrected OAM spectrums. (a), (c), (e), (g), (i), (k) and (m) shows the measured OAM spectrums of beams carrying OAM of $l$=-3~3, respectively, while (b), (d), (f), (h), (j), (l) and (n) shows the corresponding corrected OAM spectrums. In figures of measured OAM spectrums, the tilt angle $\theta$ and dispersion factor $\upsilon$ are labeled while in figures of corrected OAM spectrums, only the dispersion factor $\upsilon$ are labeled.

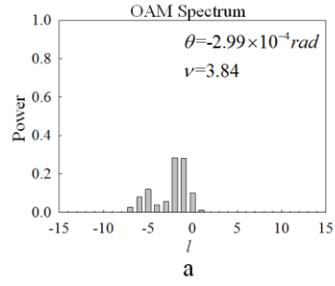
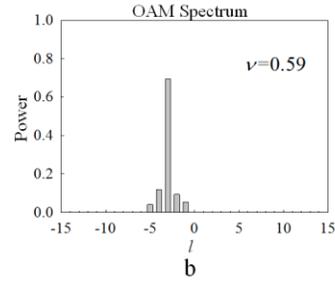

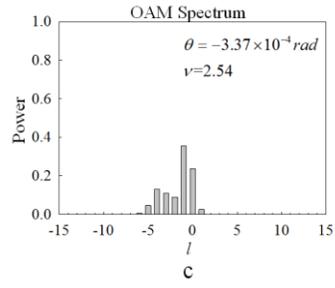
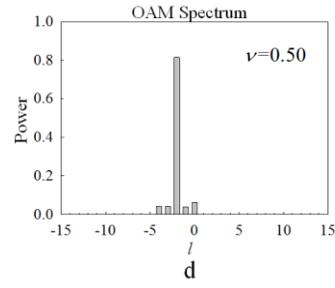

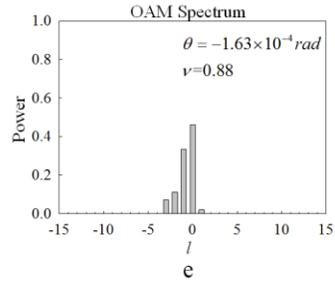
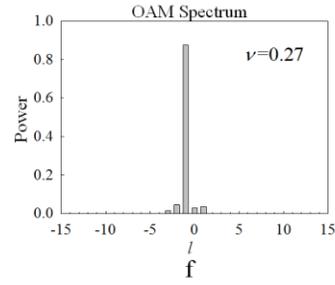

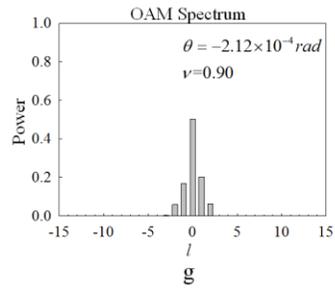
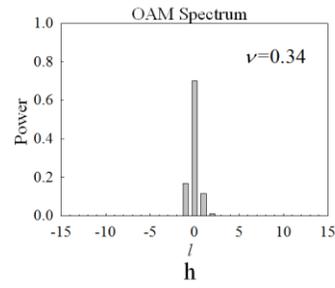

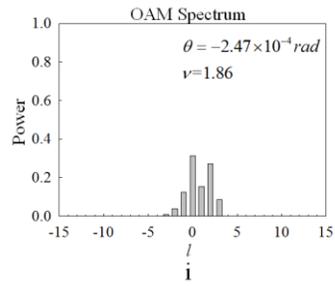
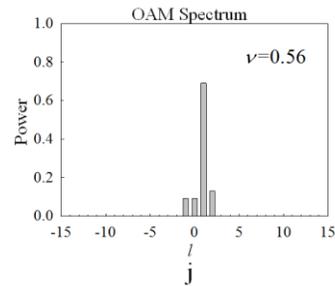

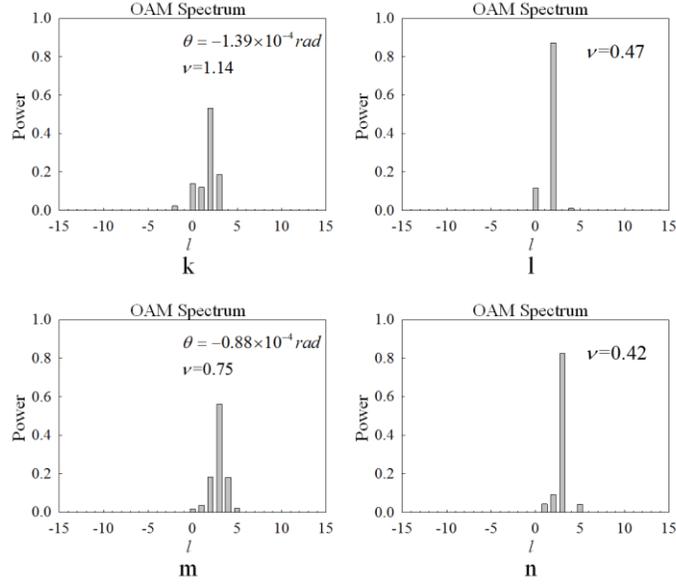

Fig. 3. The measured OAM spectrums and the corrected OAM spectrums. (a), (c), (e), (g), (i), (k) and (m) shows the measured OAM spectrums of beams carrying OAM of $l=-3\sim3$, respectively, while (b), (d), (f), (h), (j), (l) and (n) shows the corresponding corrected OAM spectrums.

**Superposition OAM beams:**
Figure 4(a) shows the measured OAM spectrum of beam superposed by OAM of $l=3$ and 0 with tilt angle of -$4.60\times10^{-4}rad$, while (c) shows the measured OAM spectrum of beam superposed by OAM of $l=-3$ and 3 with tilt angle of $-4.15\times10^{-4}rad$. Fig. 4(b) and (d) show the corrected OAM spectrum corresponding to (a) and (c), respectively.

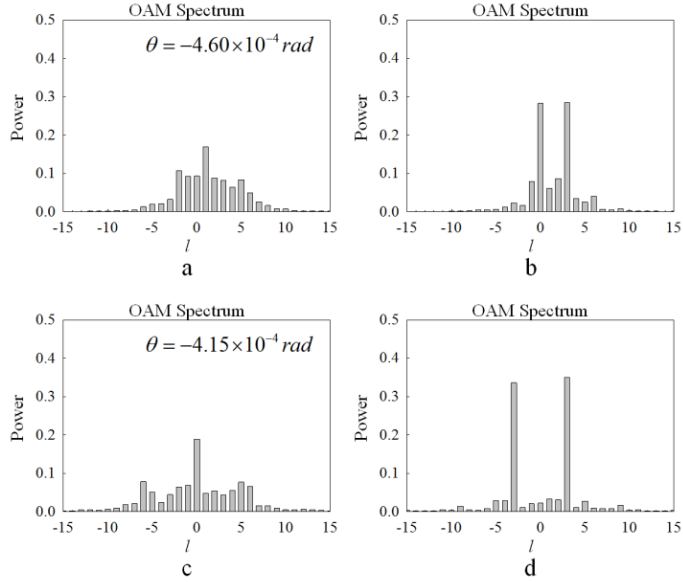

Fig. 4. (a) shows the measured OAM spectrum of beam superposed by OAM of $l=3$ and 0 with tilt angle of -$4.60\times10^{-4}rad$, while (c) shows the measured OAM spectrum of beam superposed by OAM of $l=-3$ and 3 with tilt angle of $-4.15\times10^{-4}rad$. (b) and (d) show the corrected OAM spectrum corresponding to (a) and (c), respectively.